\documentclass[twocolumn,twoside]{IEEEtran}

\usepackage{graphicx}
\usepackage{wrapfig}
\usepackage{epsfig}
\usepackage{graphicx}
\usepackage{color}
\usepackage{mdframed}
\usepackage{mathrsfs} %\usepackage{epic}
\usepackage{amsfonts}
\usepackage{latexsym}
\usepackage{amsmath}
\usepackage{amssymb}
\usepackage{color}
\usepackage{float} 
\usepackage{floatflt} 
\usepackage{caption}
\usepackage{subcaption}
\usepackage{balance}
\usepackage{cite}
\usepackage{pdfpages}

\newcommand{\suppress}[1]{}

\newtheorem{theorem}{Theorem}[section]
\newtheorem{question}{Question}%[section]

\newtheorem{claim}{Claim}[section]

\newtheorem{definition}{Definition}[section]
\newtheorem{corollary}{Corollary}[section]
\newtheorem{remark}{Remark}[section]

\newcommand{\bRk}{{\bf R}_{\tt key}}
\newcommand{\bRkk}{{\bf R}_{\tt key(2)}}

\newcommand{\bRs}{{\bf R}_{\tt sec}}

\newcommand{\rs}{{\tt sec}}
\newcommand{\rk}{{\tt key}}

\def\proof{\par\penalty-1000\vskip .5 pt\noindent{\bf Proof\/: }}

\def\proof{\noindent\hspace{2em}{\it Proof: }}

\def\cA{\mbox{$\cal{A}$}}

\def\cX{\mbox{$\cal{X}$}}

\def\cB{\mbox{$\cal{B}$}}

\newcommand{\cI}{{{\cal I}}}

\def\01{\{0,1\}}

\newcommand{\remove}[1]{}

\begin{document}

%\thispagestyle{empty}
%\IEEEoverridecommandlockouts
\title{Network Coding Multicast Key-Capacity}
%Secret Key Multicast via Network Coding

\author{Michael Langberg\ \ \ \ \ \ \ \ \  \ \ \ \ \ \ \ \ \ 
%~\IEEEmembership{Fellow,~IEEE},
Michelle Effros
%,~\IEEEmembership{Fellow,~IEEE} \\
%M. F. Wong \IEEEmembership{Student Member,~IEEE}
%\thanks{Manuscript received January xx, 2013.}
\thanks{M. Langberg is with the Department of Electrical Engineering at the University at Buffalo (State University of New York).  
Email: \texttt{mikel@buffalo.edu}}
\thanks{M. Effros is with the Department of Electrical Engineering at the California Institute of Technology.
Email: \texttt{effros@caltech.edu}}
%\thanks{M. F. Wong is with the Department of Electrical Engineering at the California Institute of Technology.
%Email : \texttt{***}}
\thanks{This work is supported in part by NSF grants CCF-1817241 and CCF-1909451. 
%The full version of this work appears in \cite{full_version}.
}
%{\tt https://arxiv.org/abs/2005.10315}}
%\thanks{A preliminary version of this work was presented in \cite{LE11,LE:12}.}
}

%\author{Michael Langberg
%\thanks{M. Langberg is with the Department of Electrical Engineering at The State University of New York at Buffalo.  
%Email : \texttt{mikel@buffalo.edu}}\ \ \ \ \ \ \ \ \  \ \ \ \ \ \ \ \ \ 
%%~\IEEEmembership{Senior Member,~IEEE},
%Michelle Effros
%%,~\IEEEmembership{Fellow,~IEEE} \\
%%M. F. Wong \IEEEmembership{Student Member,~IEEE}
%%\thanks{Manuscript received January xx, 2013.}
%\thanks{M. Effros is with the Department of Electrical Engineering at the California Institute of Technology.
%Email : \texttt{effros@caltech.edu}}
%%\thanks{M. F. Wong is with the Department of Electrical Engineering at the California Institute of Technology.
%%Email : \texttt{***}}
%\thanks{This material is based upon work supported by NSF grants CCF-***.}
%%\thanks{A preliminary version of this work was presented in \cite{LE11,LE:12}.}
%}

\maketitle

\begin{abstract}
For a multi-source multi-terminal noiseless network, the {\em key-dissemination} problem involves the task of multicasting a secret key $K$ from the network sources to its terminals. 
As in secure multicast network-coding, in the key-dissemination problem the source nodes have access to independent randomness and, as the network is noiseless, the resulting key $K$ is a function of the sources' information.
However, different from traditional forms of multicast, in key-dissemination the key $K$ need not  consist of source messages, but rather may be {\em any} function of the information generated at the sources, as long as it is shared by all terminals.
Allowing the shared key $K$ to be a mixture of source information grants a flexibility to the communication process which gives rise to the potential of increased key-rates when compared to traditional secure multicast. 
The multicast {\em key-capacity} is the supremum of achievable key-rates, subject to the security requirement that the shared key is not revealed to an eavesdropper with predefined eavesdropping capabilities.
The key-dissemination problem (termed also, secret key-agreement) has seen significant studies over the past decades in memoryless network structures.
In this work, we initiate the study of key-dissemination in the context of noiseless networks, i.e., network coding.
In this context, we study similarities and differences between traditional secure-multicast and the more lenient task of key-dissemination. 
\end{abstract}

\section{Introduction} 
\label{sec:intro}

A {\em key-dissemination} communication protocol is one in which a key $K$, which is  at times secret, is shared among a collection of users as a prelude to future communication tasks requiring shared user common knowledge. The task of key dissemination (termed also, secret key-agreement) has seen significant studies over the past decades in memoryless network structures, e.g., \cite{wyner1975wire,csiszar1978broadcast,ahlswede1993common,maurer1993secret,csiszar2004secrecy,chan2014multiterminal,csiszar2008secrecy,gohari2010information,gohari2010information2,siavoshani2010group,xu2016private,hayashi2016secret,narayan2016multiterminal} in which a collection of nodes wish to share a common key over a noisy network structure which is subject to eavesdropping. 
Typical network structures in the studies above include a broadcast channel enhanced with a public noiseless-channel, where the key is generated at the source node and the eavesdropper has both noisy access to the broadcasted information and noiseless access to the public channel.
Remarkably, the public channel improves on the achievable key rate despite being completely exposed to eavesdropping. 
%\ml{Maybe give more details}

This work initiates the study of key-dissemination in the context of noiseless networks, i.e., in the context of Network Coding.
Roughly speaking, for a multi-source, multi-terminal network, in the key-dissemination problem one wishes to multicast a key $K$ of rate $R$ from a collection of sources to a collection of terminal nodes. 
Sources have access to independent randomness, and, as the network is noiseless, the resulting key $K$ is a function of the sources' information.
However, unlike traditional forms of secure multicast, there is no requirement on $K$ beyond the following three constraints. First, $K$ should be delivered to all terminal nodes, second, $K$ should not be revealed to an eavesdropper with predefined eavesdropping capabilities, and third, $K$ is uniform and has rate at least $R$. 
Allowing the shared key $K$ to be any function of the source information grants a flexibility to the communication process which gives rise to the potential of increased key-rates when compared to traditional secure multicast. 
Given a network instance, one seeks to determine the {\em key-capacity}, naturally defined  
as the closure of all achievable key rates.   Formal definitions of the concepts above (and additional ones that appear below) are given in detail in Section~\ref{sec:model}.

The eavesdropper capabilities in the model under study may differ depending on the motivation at hand. For example, one may consider the extreme scenario in which every network node is considered a malicious entity; here, we require that for each non-terminal network node $v$, {\em including} each source node contributing to the randomness determining $K$,  the information passing through $v$ is independent of $K$. 
Figure~\ref{fig:examples}.a depicts an example.
%Here, one requires that even the source nodes themselves, that generate the randomness that eventually determines the key $K$, gain no-information regarding $K$.
This scenario may be appropriate for network protocols that share a secret key between a pair of users  to later be used as a one-time-pad for the secure communication of sensitive information.

On the other extreme, consider a setting in which no security is required.
Such a setting may be applicable for communication among trusted parties.
Now, key-dissemination  becomes the task of communicating {\em any form} of shared information $K$ to the network terminals without constraining what other network components may learn.
While this setting resembles that of traditional (non-secure) communication, it leaves open the possibility of increased rate due to the possibility that $K$ may be any function of the sources but need not be sufficient to reconstruct source information.

\subsection{Related work}
\label{sec:related}

The problem of network-coding, multi-source, multi-terminal, key-dissemination is closely related to the task of {\em secure ``wiretap'' multicast network coding} in which the goal is to securely multicast source information to a collection of terminals in the presence of an eavesdropper with, as above, predefined eavesdropping capabilities.
In full generality, the model of secure multicast network-coding distinguishes between source nodes that have access to message information, and nodes that generate independent randomness used to enable secure communication.
The majority of prior works study single-source multicast in which the single source node generates both messages and independent randomness, i.e., no additional network nodes can generate randomness, and the eavesdropper can access any collection of up-to $z$ network links for a security parameter $z$, e.g., 
\cite{cai2002secure,feldman2004capacity,cai2007security,yeung2008optimality,el2012secure,silva2011universal,jaggi2012secure}. A major result in this context includes a characterization of the secure multicast capacity, which can be efficiently obtained by linear codes.

The model in full generality, where several network nodes may generate messages and/or independent randomness, is studied in, e.g., \cite{huang2018,chan2014network, cui2012secure, chaudhuri2018trade, chaudhuri2019secure,chaudhuri2021characterization};
%chaudhuri2019secure, chaudhuri2018trade, 
%ISIT**
its capacity is less well understood. 
Specifically,  \cite{cui2012secure} shows that determining the secure-multicast capacity in  instances with a 
single message-generating source, a single terminal, and certain eavesdropping capabilities is NP-Hard;   \cite{huang2018,chan2014network} show, for single-source, single-terminal settings in which the eavesdropper can access any single ($z=1$) edge in the network (each of unit capacity), that determining the secure-rate when any node can generate random keys is as hard as the problem of characterizing the (non-secure) capacity region of the $k$-unicast problem. 
The $k$-unicast problem is a well known open problem in the study of network codes, e.g., \cite{4460828,chan2014network,6293890,langberg2009multiple}. 
%jalali2012capacity,,4460828
%ISIT**

Secure network coding and key-dissemination are similar in the sense that the information eventually shared between terminals is kept secret from the network eavesdropper.
They differ in that in the former source nodes hold message information that must be recovered while in the latter the key $K$ may be any function of the independent randomness held by the source nodes.
The flexible decoding in key-dissemination opens the possibility of a key-rate $R$ that exceeds the secure-multicast capacity.
The study at hand addresses the differences and similarities between key-dissemination and secure-multicast. 
Our results are summarized in Section~\ref{sec:questions}.

%The problem of key-dissemination is related to other tasks beyond that of secure multicast.
%%Examples, discussed in greater detail in the full version of this work \cite{LE:22},  include network coding scenarios in which the communicated information is something other than pure source bits. 
%Examples, discussed in greater detail in the full version of this work \cite{LE:22}, include studies in network coding {\em function computation}, e.g., \cite{appuswamy2011network,kowshik2012optimal,shah2013network}, in which a predetermined function of the source information, such as a sum of source values \cite{ramamoorthy2013communicating,rai2012network,shenvi2010necessary,appuswamy2013computing,li2022arithmetic}, is to be shared between all terminal nodes.
%{\em Pliable index coding}, e.g., \cite{brahma2015pliable,liu2019private,sasi2019code,liu2020secure} is an example prior work in which the information decoded at terminals is of a flexible nature loosely reminiscent of the flexibility of the key $K$ in the key-dissemination problem.
%Finally, the problem of secret-key generation in the context of wireless networks using the methodology of network coding (i.e., that of performing coding operations at internal network nodes) has appeared, for example, in \cite{oliveira2008network,xiao2018cooperative,lima2009towards}.
%The models, questions, and results of the works above differ significantly from those presented in this work.

%ISIT**
%****
%

\begin{figure}[t!]
\hspace{6mm}
\includegraphics[scale=0.4]{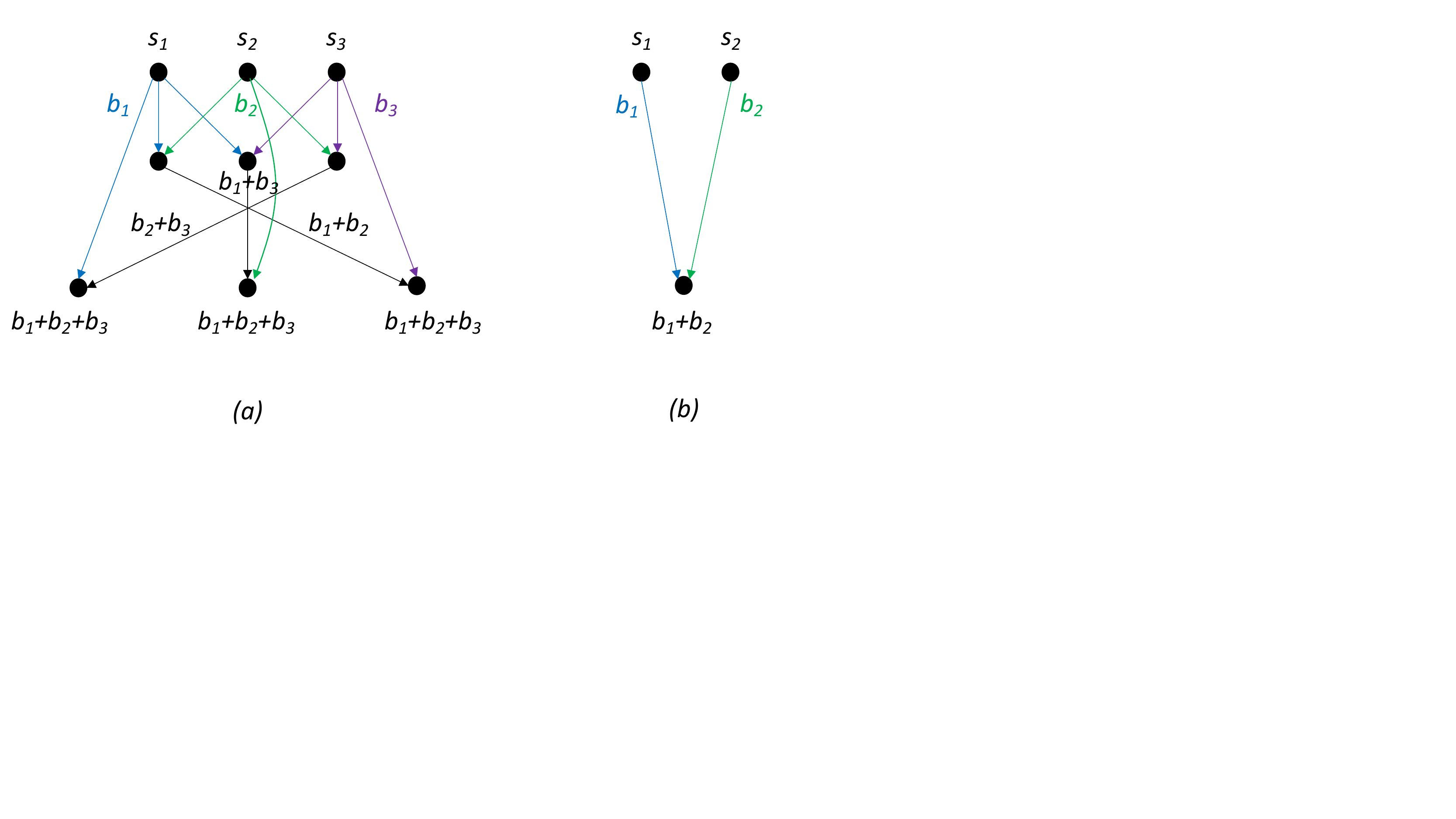}
\vspace{-40mm}
\caption{{\small Two simple example networks expressing the difference between secure-multicast and key-dissemination. Source $s_i$ generates random bit $b_i$. Terminals appear on the lowest layer of the networks. All edges are of capacity 1. Networks {\em (a)} and {\em (b)} are examples in which the key-dissemination capacity is 1, with $K$ being the sum-of-sources, even when the eavesdropper is capable of accessing all information available to any single non-terminal node of the network (including the source nodes).
The secure-multicast capacity with such an eavesdropper is 0. While Network {\em (b)} is a trivial such example, Network {\em (a)} also acts as a simplified example for the proof of Theorem~\ref{thm:gap} exhibiting a multiplicative advantage of $\alpha=2$ to answer Question~\ref{q:mix} {\em ``To mix or not to mix,''} with a definitive demonstration that mixing offers a rate advantage (giving $\bRk(\cI)=1$ and $\bRkk(\cI)=0.5$).}}
%, and a non-secure multicast sum-rate of at most 1.5).}}
\label{fig:examples}
\end{figure}

The problem of key-dissemination is related to other tasks beyond that of secure multicast.
Examples include network coding scenarios in which the communicated information is something other that pure source bits.
For example, network coding function-communication, e.g., \cite{appuswamy2011network,kowshik2012optimal,shah2013network}, in which a predetermined function of the source information, such as a sum of source values \cite{ramamoorthy2013communicating,rai2012network,shenvi2010necessary,appuswamy2013computing,li2022arithmetic}, is to be shared between all terminal nodes.
Sum function network codes might lend themselves to the problem of key-dissemination, as a key $K$ set to be the sum of all source information is independent of partial sums (including individual source randomness) communicated over network edges. 
Two networks illustrating key dissemination using sum function network codes are depicted in Figure~\ref{fig:examples}.
As with secure multicast in it general form, determining the capacity of sum networks is as hard as determining the capacity of multiple-unicast network coding \cite{rai2012network}; this is shown through a reduction implying, rather counter intuitively, that linear codes do not suffice to achieve capacity in sum-networks.

In both secure-multicast and functional-communication, the information transmitted is required to be a certain predetermined function of the source information. 
{\em Pliable index coding} \cite{brahma2015pliable} is an example prior work in which the information decoded at terminals is of a flexible nature loosely reminiscent of the flexibility of the key $K$ in the key-dissemination problem.
Index coding is a representative form of multiple-unicast network coding \cite{bar2011index,el2010index,effros2015equivalence} in which a server holding all source messages wishes to communicate through a capacity-limited noiseless broadcast channel with multiple terminals, each holding potentially distinct message side-information and requiring potentially distinct messages.
Pliable index coding \cite{brahma2015pliable} is a variant of index coding in which terminals are required to decode not a specific source message, but {\em any} message they do not already have as side information, a flexibility implying significant rate advantages when compared to traditional index coding.
Various forms of security in the context of  index coding and pliable index-coding have been studied, e.g., in \cite{dau2012security,liu2019private,sasi2019code,liu2020secure}.

Finally, the problem of secret-key generation in the context of wireless networks using the methodology of network coding (i.e., that of performing coding operations at internal network nodes) has appeared, for example, in the context of sensor networks \cite{oliveira2008network}, dynamic wireless systems \cite{xiao2018cooperative}, and multiresolutional streaming \cite{lima2009towards}.
The models, questions, and results of the works above differ significantly from those presented in this work.

\subsection{Main questions and results addressed in this work}
\label{sec:questions}

In this work we study the relationship between the key-dissemination problem and the traditional secure-multicast problem.
Our study is guided by the following two questions:

\begin{question}
\label{q:mix}
{\bf To mix or not to mix?} {\em Does the flexibility allowing $K$ to be {\em any} function of source randomness improve the rate of communication when compared to traditional communication, in which pre-specified messages must be decoded at terminals. In other words, does allowing terminals the ability to directly decode a {\em mixture} of source randomness hold rate benefits?}
\end{question}

\begin{question}
\label{q:complexity}
{\bf How hard is key-dissemination?} {\em While efficient, capacity-achieving codes for non-secure (multiple-source) multicast network coding and for certain settings of secure multicast are well understood, and the design of such codes for other settings of secure multicast is currently open, what can be said for the key-dissemination problem regarding the tasks of determining the capacity and mastering code-design? }
\end{question}

In our study of the key-dissemination problem, we present the following results. Our results are presented below in a loose manner and stated rigorously after our model is presented in Section~\ref{sec:model}.

\subsubsection{Single-source case}
When only one source can generate randomness, we show that key-dissemination is equivalent to secure multicast. Namely, with respect to Question~\ref{q:mix}, there is no benefit in this setting to solutions that ``mix,'' i.e., to solutions in which terminals directly decode a mixture of source randomness. Moreover, with respect to Question~\ref{q:complexity}, code design and capacity are well understood for certain eavesdropping capabilities and are open for others; this corresponds to the state of the art for secure multicast. Our results for the single-source case are presented in Theorem~\ref{thm:single}.

%\subsubsection{Single-terminal case} ***
%When only one source can generate randomness, it is fairly straight-forward to show that key-dissemination reduces to secure multicast. Namely, with respect to Question~\ref{q:mix}, there is no benefit in this setting to ``mix'', i.e., to allow terminals the ability to directly decode a mixture of source randomness. Moreover, with respect to Question~\ref{q:complexity}, code design and capacity are well understood for certain eavesdropping capabilities, and are open for others (as is the state-of-the-art for secure multicast)

\subsubsection{Non-secure case} 
In the non-secure multi-source setting of key-dissemination, one wishes to establish shared randomness $K$ among terminal nodes, but does not need to protect $K$ from other network components.
While this setting resembles that of traditional (non-secure) multicast, it leaves open the possibility that directly decoding a ``mixture'' of source randomness (in the sense of Question~\ref{q:mix}) may increase the key rate. 
For linear codes, which are capacity achieving for traditional (non-secure) multi-source multicast network coding, we show in Theorem~\ref{thm:Bzero} that mixing does not help, thus resolving Question~\ref{q:mix} in the negative.  
For general codes in the non-secure case,  Questions~\ref{q:mix} and ~\ref{q:complexity} remain open and are subject to future work.

\subsubsection{General case, complexity} 
We study Question~\ref{q:complexity} in the context of key dissemination and  show in Theorem~\ref{thm:hard_k_s} and Corollary~\ref{cor:hard} that computing the capacity of the key dissemination problem even when only single edges may be eavesdropped, is as hard as determining the multiple-unicast network coding capacity. 
Our hardness result is based on reducing hard instances of secure multicast to the key-dissemination problem. 
%Connections in the opposite direction (reducing key-dissemination to secure multicast) are also shown to exist in \ml{Theorem~\ref{thm:s_k}.}

\subsubsection{General case, mixing} 
Finally, we study Question~\ref{q:mix} in the context of key dissemination in its general form.
For multi-source key-dissemination, depending on the eavesdropping capabilities, it is not hard to construct simple instances that have large key-dissemination rate while the corresponding secure-multicast rate is zero. 
%Thus answering Question~\ref{q:mix} affirmatively. 
As a result, mixing has an advantage here.
Two such instances are given in Figure~\ref{fig:examples}.
In each we assume that the eavesdropper has access to source nodes, and therefor it is necessary for each source random variable to be independent from the shared key $K$. 
Such a stringent security requirement does not allow a positive secure-multicast rate, but  mixing at terminal nodes may allow a large key-dissemination rate.

To better understand the potential benefits in allowing the decoders to {\em directly} decode a mixture of source randomness, we compare between two potential decoding procedures in the context of key-dissemination.
First consider a 2-stage decoding procedure in which each terminal starts by decoding source information (as in the setting of secure-multicast) and only then proceeds in defining $K$ to be a function of the decoded source information from the first stage.
For example, as in Figure~\ref{fig:examples}.b, one may consider a terminal that decodes, in the first stage, message $b_1$ from source 1 and $b_2$ from source 2, and defines the key $K$ to be the sum $b_1+b_2$ in the second stage. The information held by each source is not independent of the decoded information $(b_1,b_2)$ of the first stage, but is independent of the final key $K$. 
Implying that, while the secure-multicast rate in this case is zero, one can still obtain a key $K$ from first decoding source messages and then combining them in a secure way to form  $K$.
Such 2-stage decoders (defined in detail in Section~\ref{sec:results}) are natural for key dissemination; however, they  may still be inferior when compared to unrestricted decoders that can directly decode $K$ from their incoming information.
Indeed, in Theorem~\ref{thm:gap}, we show instances for which there is a multiplicative benefit in rate (that grows with the network size) to directly decoding $K$ over the 2-stage decoder.
One such example network, exhibiting a multiplicative benefit of 2, is depicted in Figure~\ref{fig:examples}.a.

%we show in mixture of source randomness
%each  source messages are not independent of the
%of we present multi-source multi-terminal instances of key-dissemination which exhibit an arbitrarily large gap between the rate achievable with and without allowing the decoders to directly decode a ``mixture'' of source randomness.
%One such instance, in which every network node may be eavesdropped, 
%that exhibits a multiplicative gap fo size \ml{Add gap}, is given in Figure~\ref{fig:examples}.
%Unlimited gaps are presented in Theorem~\ref{thm:gap}.
%For Question~\ref{q:complexity}
%
%to study the benefit in allowing the decoders to directly decode a ``mixture'' of source randomness,
%we compare the rate achievable in two scenario ...
%
%relax Question~\ref{q:mix} in this case, and ask if one performs
%
%***

%we present multi-source multi-terminal instances of key-dissemination which exhibit an arbitrarily large gap between the rate achievable with and without allowing the decoders to directly decode a ``mixture'' of source randomness.
%One such instance, in which every network node may be eavesdropped, 
%that exhibits a multiplicative gap fo size \ml{Add gap}, is given in Figure~\ref{fig:examples}.
%Unlimited gaps are presented in Theorem~\ref{thm:gap}.
%For Question~\ref{q:complexity}

\section{Model} 
\label{sec:model}
%Throughout the paper, the size of a finite set $S$ is denoted by $|S|$. 
%For any positive real $k$, $[k]$ denotes the set $\{1,...,\lfloor k \rfloor\}$.
%We use bold letters to denote vectors; for example, $\bR=(R_1,...,R_{k})$ is a vector of dimension $k$ and $R_i$ is the $i^{th}$ element of vector $\bR$.
%We define $\mathbf{R}-\gamma$ as $((R_1-\gamma)^+,...,(R_{k}-\gamma)^+)$ where
%%We use $\oplus$ to denote binary addition.
%$(R-\gamma)^+ = \max\{0,R-\gamma\}$.
%For $\alpha>0$ and a set $\cR$ of real vectors, the set $\alpha \cR$ refers to the set obtained by multiplying each vector in $\cR$ by $\alpha$.

\subsection{Multicast Network Coding}
{\bf Network Coding Instance:} An instance ${\mathcal I}=(G,S,D,\cB)$ of the network coding problem includes an acyclic\footnote{We assume acyclicity for simplicity.  Using standard techniques outlined, e.g.,  in \cite{langberg2021edge}, our results hold also for cyclic networks.} directed network $G=(V,E)$ in which each edge $e \in E$ has an associated capacity $c_e$, a 
collection of source nodes $S \subseteq V$, a collection of terminal nodes $D \subseteq V$, and a collection of subsets of edges $\cB=\{\beta_1,\dots,\beta_{{\tiny |\cB|}}\}$, $\beta_i \subseteq E$ that may be subject to eavesdropping.

Each source node $s_i \in S$ holds an unlimited collection of independent, uniformly distributed bits $\{b_{ij}\}_j$.  
Given the acyclic nature of $G$, we assume that communication occurs according to the topological order in $V$, where for blocklength $n$ every edge $e \in E$ carries a message over an alphabet $\cX^n_e$ of size $\lfloor 2^{c_e n}\rfloor$. 
Roughly speaking, multicast-communication at rate $R$ is successful if at the end of the communication process all terminals $d \in D$ share a random variable $K$ uniformly distributed over $[2^{Rn}]$ that is independent from the information held by any individual subset of edges $\beta \in \cB$.
Here, for $x>0$, $[x]=\{1,2,\dots,\lfloor x \rfloor\}$.
%For traditional secure network coding, $K$ is required to consist solely of bits $b_{ij}$ held by sources in $S$. In the key-dissemination problem, $K$ is not restricted in any way and be any function of the source information $\{b_{ij}\}_{ij}$.
\vspace{2mm}

\noindent
{\bf Network Codes:}
More formally, for blocklegth $n$, network code 
$({\mathcal F},\mathcal{G})=(\{f_{e}\},\{g_j\})$
is an assignment of encoding functions 
$\{f_{e}\}$ for each edge $e\in E$ and a decoding function $g_j$ to each terminal $d_j\in D$. 
For every edge $e=(u,v)$, the edge message
$X^n_{e} \in \cX^n_{e}$ from $u$ to $v$ is equal to the evaluation of encoding function $f_{e}$  on inputs $X^n_{{\rm In}(u)}$. 
Here, for a generic node $u_0$, $X^n_{{\rm In}(u_0)}$ equals
$((X^n_{e'}:e' = (v,u_0) \in E), (\{b_{ij}\}_j: u_0=s_i))$ 
and captures all information available to node $u_0$ during the communication process.
Communication proceeds according to a topological order on $E$ and is considered successful if for every terminal $d_j \in D$ the evaluation of decoding functions $g_{j}$ on the vector of random variables $X^n_{{\rm In}(d_j)}$ equals the reproduction of a uniform random variable $K$ over alphabet $[2^{Rn}]$ for a target rate $R$ such that for every $\beta \in B$, $I(K;(X^n_e: e \in \beta))=0$. 
That is, we seek zero-error key-dissemination with perfect security\footnote{Although we do not discuss asymptotically vanishing error and/or weaker security requirements in this work, our results can be extended to these settings given the broad nature of the results in \cite{huang2018}.}.
Specifically,
%Below, we distinguish between feasibility for key-distribution and feasibility for traditional secure communication. 
\vspace{2mm}

\noindent
{\bf Key-dissemination feasibility:} Instance $\cI$ is said to be $(R,n)_\rk$-feasible if there exists a network code $({\mathcal F},\mathcal{G})$ with blocklength $n$ such that
%with probability at least $1-\e$, 
\begin{itemize}
	\item {\bf Key Rate:} $K$ is a uniform random variable with $H(K)=Rn$.
	\item {\bf Decoding:} For all $d_j \in D$, $H(K|X^n_{{\rm In}(d_j)})=0$.
	\item {\bf Secrecy:}  $I(K;(X^n_{e}:e \in \beta))=0$ for any subset $\beta \in \cB$.
\end{itemize}

\vspace{2mm}

\noindent
{\bf Secure-multicast (sum-rate) feasibility:} Our model slightly changes when discussing secure-multicast. In the secure-multicast setting, one distinguishes between source-nodes $S_m$ that hold message information and source nodes $S_r$ that hold independent randomness used for masking. The two subsets may intersect. As before, we assume that every node $s_i$ in $S_m \cup S_r$ holds an unlimited collection of independent bits $\{b_{ij}\}_j$. Instance $\cI=(G,(S_m,S_r),D,\cB)$ is said to be $(R,n)_\rs$-feasible if there exists a network code $({\mathcal F},\mathcal{G})$ with blocklength $n$ such that
%with probability at least $1-\e$, 
\begin{itemize}
	\item {\bf Message Rate:} $K$ is a uniform random variable with $H(K)=Rn$ such that $K$ equals a collection of bits included in $(b_{ij}: s_i \in S_m)$, i.e., bits generated by sources in $S_m$.
	\item {\bf Decoding:} For all $d_j \in D$, $H(K|X^n_{{\rm In}(d_j)})=0$.
	\item {\bf Secrecy:}  $I(K;(X^n_{e}:e \in \beta))=0$ for any subset $\beta \in \cB$.
\end{itemize}

Notice the difference between secure-multicast feasibility and key-dissemination feasibility, in the former the key $K$ consists of a collection of random bits $b_{ij}$ generated at source nodes $s_i \in S_m$ while in the latter $K$ may consist of any function of random bits $\{b_{ij}\}_{ij}$ (of sources  $s_i \in S$).
%We distinguish between two cases. 
%In the {\bf Key Dissemination Problem}, no additional requirements are made on $K$ and we say that instance $\cI$ is $(R,n)_\rk$-feasible. 
%In the context of {\bf Traditional Secure-Communication} ones requires the key $K$ to be derived from a collection of source information bits multicasted to the receivers. Formally, $\cI$ is said to be $(R,n)_\rs$-feasible if there exists a collection of source information bits $m=(b_{ij}: (ij) \in I)$ for an index set $|I| \geq Rn$ such that 
%for each terminal $d_j$, $H(m|X_{{\rm In}(d_j)})=0$ and in addition $H(K|m)=0$.

%\begin{definition}
%	[Secure-Capacity]
%	\label{def:cap_s}
%	The multicast secure-capacity of $\cI$, denoted by $\bRs(\cI)$, is maximum $\Rs$ for which for all ${\Delta}>0$ there exist infinitely many blocklengths $n$ 
%	such that $\cI$ is $(\Rs-{\Delta},n)_\rs$-feasible. 
%Restricting all encoding and decoding operations to be linear, we define the multicast linear secure-capacity $\bRs^L(\cI)$ analogously.
%	%$\mathcal{R}^0_0(\cI)$ is the zero-error capacity region for independent sources.
%\end{definition}
 
\begin{definition}
[Key Capacity and Secure Capacity]
\label{def:cap_k}
The multicast key-capacity of $\cI$, denoted by $\bRk(\cI)$, is the maximum $R$ for which for all ${\Delta}>0$ there exist infinitely many blocklengths $n$ such that $\cI$ is $(R-{\Delta},n)_\rk$-feasible. 
%$\mathcal{R}^0_0(\cI)$ is the zero-error capacity region for independent sources.
Restricting all encoding and decoding operations to be linear, we define the multicast linear key-capacity $\bRk^L(\cI)$ analogously.
The secure capacity $\bRs(\cI)$ and its linear variant $\bRs^L(\cI)$ are the corresponding capacities.
\end{definition}

\section{Formal statement of results}
\label{sec:results}

Throughout this work we study the connections between key-dissemination and secure-multicast. 
In many of the statements below, given an instance $\cI=(G,S,D,\cB)$ of the key-dissemination problem, we define a corresponding ``refined'' instance for secure-multicast $\cI_\rs=(G,(S_m,S_r),D,\cB)$ which is identical to $\cI$ except for the definition of $S_m$ and $S_r$ which are both set to equal $S$, i.e., in $\cI_\rs$, all source nodes in $S$ can generate both message bits and random bits used for masking. 
By our definitions in Section~\ref{sec:model}, it holds for $\cI$ and the corresponding $\cI_\rs$ that  $\bRs(\cI_\rs) \leq \bRk(\cI)$, as any code that is $(R,n)_\rs$-feasible on $\cI_\rs$ is also  $(R,n)_\rk$-feasible on ${\cI}$.
Our study is motivated by the potential benefit of $\bRk(\cI)$ over $\bRs(\cI_\rs)$.

\begin{theorem}[Single source case]
\label{thm:single}
Let $\cI = (G,S,D,\cB)$ be an instance of the key-dissemination problem with $|S|=1$, and let $\cI_\rs=(G,(S_m,S_r),D,\cB)$ be the corresponding instance of the secure multicast problem with $S_m=S_r=S$, then 
$$
\bRk(\cI)=\bRs(\cI_\rs)
$$
\end{theorem}

%\begin{theorem}[Single terminal case]
%\label{thm:single_t}
%???
%Let $\cI = (G,S,D,B)$ be an instance of the network coding problem with $|S|=1$ and let $\cI_\rs$ be the corresponding instance of the secure multicast problem with $S_m=S_k=S$ then 
%$$
%\bRk(\cI)=\bRs(\cI).
%$$
%\end{theorem}

\begin{theorem}[Non-secure case]
\label{thm:Bzero}
Let $\cI = (G,S,D,\cB)$ be an instance of the key-dissemination problem with $\cB=\phi$, and let $\cI_\rs=(G,(S_m,S_r),D,\cB)$ be the corresponding instance of the secure multicast problem with $S_m=S_r=S$, then 
$$
\bRk^L(\cI)=\bRs^L(\cI_\rs).
$$
\end{theorem}

\begin{remark}
	The question of whether Theorem~\ref{thm:Bzero} holds for general (not necessarily linear) codes remains open.
	In other words, Question~\ref{q:mix} restricted to the non-secure setting, which asks if ``mixing helps,'' is unsolved.
	Equivalently, since $\bRs(\cI_\rs) = \bRs^L(\cI_\rs)$ in this case, it is unknown if there is an advantage to non-linear codes in key-dissemination when $\cB=\phi$.
%	
%	An affirmative answer would imply that $\bRk(\cI)>\bRs(\cI_\rs)$ and thus, as $\bRs(\cI_\rs) = \bRs^L(\cI_\rs)=\bRk^L(\cI)$ when $\cB=\phi$, that $\bRk(\cI) > \bRk^L(\cI)$, i.e., this would mean that there is an advantage to non-linear codes in key-dissemination when $\cB=\phi$.
\end{remark}

%For general $B$ the gap between $\bRk$ and $\bRm$ can be arbitrarily large.

\begin{theorem}
\label{thm:hard_k_s}
Let $\cI_\rs=(G,(S_m,S_r),D,\cB)$ be a secure-multicast instance with $|S_m|=1$.
%, in which 
%$G=(V,E)$, 
%$S_m=\{s\} \subseteq V$.
%, and $S_r=V$.
%and $\cB = \{\beta_e=\{e\} | e \in E\}$ consists of all single-edge subsets of $E$.
Let $R$ be a rate parameter.
One can efficiently construct an instance $\cI_\rk=(G_\rk,S_\rk,D_\rk,\cB_\rk)$  of the key dissemination problem such that 
$R \in \bRs(\cI_\rs)$ if and only if $R \in \bRk(\cI_\rk)$.
\end{theorem}

In \cite{huang2018}, it is shown that even for secure-multicast instances $\cI_\rs$ for which $S_m$ is of size 1, $D$ is of size 1, $\cB = \{\beta_e=\{e\} | e \in E\}$ consists of all single-edge subsets of $E$, all edges in $E$ are of unit capacity, and $S_r=V$, computing the secure-multicast capacity is as hard as resolving the capacity of multiple-unicast network coding instances. 
Corollary~\ref{cor:hard} follows from the instance $\cI_\rk$ obtained in the reduction from Theorem~\ref{thm:hard_k_s}.
\begin{corollary}[Key-dissemination is {\em hard}]
\label{cor:hard}
Determining the capacity of the key-dissemination problem is at least as difficult as determining the capacity of the multiple-unicast network coding problem. 
%\ml{Need to verify the exact hardness statement}
\end{corollary}

As discussed previously, to address Question~\ref{q:mix} in the general key-dissemination setting, we first define the {\em 2-stage decoding rate for key-dissemination}.
\vspace{2mm}

\noindent
\noindent
{\bf 2-stage key-dissemination feasibility:} Instance $\cI$ to the key-dissemination problem is said to be $(R,n)_{\rk(2)}$-feasible if there exists a network code $({\mathcal F},\mathcal{G})$ with blocklength $n$ such that
%with probability at least $1-\e$, 
\begin{itemize}
\item {\bf Key-rate:} $K$ is a uniform random variable with $H(K)=Rn$.
\item {\bf 2-Stage decoding:} There exists a collection $M$ of bits included in $(b_{ij}: s_i \in S)$
such that for all $d_j \in D$, $H(M|X^n_{{\rm In}(d_j)})=0$. Moreover, $K$  may be determined from $M$, i.e., $H(K|M) = 0$.
\item {\bf Secrecy:}  $I(K;(X^n_{e}:e \in \beta))=0$ for any subset $\beta \in \cB$.
\end{itemize}

%{\bf 2-stage key-dissemination feasibility:} Instance $\cI$ to the key-dissemination problem is said to be $(R,n)_{\rk(2)}$-feasible if there exists a network code $({\mathcal F},\mathcal{G})$ with blocklength $n$ such that
%%with probability at least $1-\e$, 
%\begin{itemize}
%	\item {\bf 2-Stage Key Rate:} There exists a collection of source information $M=\{b_{ij}\}_{(i,j) \in I}$ (decoded at terminals) such that the key $K$ is a uniform random variable with $H(K)=Rn$ and $K$  may be determined from $M$, i.e., $H(K|M) = 0$.
%	\item {\bf2-Stage Decoding:} For all $d_j \in D$, $H(M|X^n_{{\rm In}(d_j)})=0$.
%
%	
%	\item {\bf Secrecy:}  $I(K;(X^n_{e}:e \in \beta))=0$ for any subset $\beta \in \cB$.
%\end{itemize}

The 2-stage key-dissemination capacity $\bRkk(\cI)$ of instance $\cI$ is defined analogously to the key-capacity $\bRk(\cI)$ of Definition~\ref{def:cap_k}.

We are now ready to state our theorem comparing $\bRkk(\cI)$ with  $\bRk(\cI)$.

\begin{theorem}[General case, mixing helps]
	\label{thm:gap}
For any integer $\alpha>1$, there exist instances $\cI = (G,S,D,\cB)$ of the key-dissemination problem such that   
$$
\bRk(\cI) \ge \alpha\bRkk(\cI)
$$
\end{theorem}

%\section{Upper and lower bounds for $\bRk$}
%\begin{itemize}
%
%\item Present ``natural scheme'' lower bound based on LP based solution.
%\item Present upper bounds based on $\bRm$? Other upper bounds?
%\end{itemize}

\section{Proofs}

%ISIT**
Before presenting our proofs, we here roughly outline the proof ideas.
In the single source case of Theorem~\ref{thm:single}, any uniform key $K$ obtained through key dissemination (potentially via mixing operations at the terminal nodes in the sense of Question~\ref{q:mix}) can be replaced by a collection of message bits, as required in secure-multicast, using an appropriate pre-encoding function at the single source. 
In the non-secure case of Theorem~\ref{thm:Bzero}, any uniform key $K$ obtained through (linear) key dissemination can be replaced by a collection of message bits across different sources through an iterative process in which, at each step, an identified bit $b_{ij}$ (held by some source $s_i$) that is independent of $K$ is deterministically set to 0. This process reduces the support of $K$ and can be shown to preserve key rate. One proceeds until $K$ can be represented as a collection of message bits as required in secure-multicast.
The reduction in Theorem~\ref{thm:hard_k_s} essentially uses an identical instance $\cI_\rk \simeq \cI_{\rs}$, with the requirement that in $\cI_\rk$ any shared key $K$ is a function of information generated at $S_m$ corresponding to message-bits in $\cI_\rs$. This is obtained by adding to $\cI_\rk$ an additional terminal that is only connected from $S_m$.
Finally, the proof of Theorem~\ref{thm:gap} involves instances reminiscent of {\em combination networks} \cite{ngai2004network}, that, on one hand, allow a key capacity of 1 by multicasting a key $K$ equal to the sum-of-sources, and, on the other, are designed to have a limited non-secure multicast sum-rate. The later, together with the pre-defined security requirements, limits the 2-stage key-capacity to obtain the stated gap. 
A simplified example network is depicted in Figure~\ref{fig:examples}.a for the special case of $\alpha = 2$.

\subsection*{{\bf Proof of Theorem~\ref{thm:single}:}
Let $\cI = (G,S,D,\cB)$ be an instance of the key-dissemination problem with $|S|=1$, and let $\cI_\rs=(G,(S_m,S_r),D,\cB)$ be the corresponding instance of the secure multicast problem with $S_m=S_r=S$, then 
$$
\bRk(\cI)=\bRs(\cI_\rs)
$$
}

\proof
The fact that $\bRk(\cI) \geq \bRs(\cI_\rs)$ follows from our definitions as discussed above. To prove that $\bRk(\cI) \leq \bRs(\cI_\rs)$, consider a network  code $({\mathcal F},\mathcal{G})=(\{f_{e}\},\{g_j\})$ for $\cI$  that is $(R,n)_\rk$-feasible. 
Let $K=f(m)$ where $f$ is the global-encoding function for $K$ and  $m=(b_{j}: j \in [\ell])$ is the vector of random bits used by the (single) source $s$ in the communication over $\cI$.
Here, for an integer $\ell$, we denote the set $\{1,2,\dots,\ell\}$ by $[\ell]$.
If $|m|=\ell=Rn$, then by our definitions it follows that $K$ must equal a permutation of $m$;  thus slightly modifying the decoding functions in $\cI_\rs$ to output $m$ we obtain an $(R,n)_\rs$-feasible  code for $\cI_\rs$.

Let $|m|=\ell>Rn$. As $K$ is uniform, for each instance $k$ of $K$ the preimage $f^{-1}(k)$ has size exactly $2^{\ell-Rn}$. Thus, there exists a pre-encoding permutation $\pi$ over $\{0,1\}^{\ell}$ for which for all $k$, $\pi^{-1}(f^{-1}(k))$ is of size exactly $2^{\ell-Rn}$ and the mapping $f(\pi(m))$ depends only on $m'= (b_{j}:  j \in [Rn])$. 
This implies that the code that first uses the pre-encoding $\pi$ on $m$ and then proceeds using  $({\mathcal F},\mathcal{G})$ is $(R,n)_\rs$-feasible. 
Specifically,
\begin{itemize}
	\item {\bf Message Rate:} $K$ is a uniform random variable with $H(K)=Rn$ such that $K$ equals the collection of bits $(b_{j}: j \in [Rn])$ generated by the single source $s \in S_m=S_r=S$.
	\item {\bf Decoding:} For all $d_j \in D$, $H(K|X^n_{{\rm In}(d_j)})=0$.
	\item {\bf Secrecy:}  Let $\beta \in \cB$, and let $h_\beta(m)$ represent the global encoding function of the original code $({\mathcal F},\mathcal{G})$ for $\cI$ corresponding to $(X_{e}: e \in \beta)$. In the original code, we have, for any $\beta \in \cB$, that $I(K;(X_{e}: e \in \beta))=I(f(m);h_\beta(m))=0.$ 
	In the new code for $\cI_\rs$, the edges $e \in \beta$ transmit $h_\beta(\pi(m))$. As $\pi$ is a permutation on $\{0,1\}^{\ell}$ and  $m$ is uniform, it now follows  in the new code that  $I(K;h_\beta(\pi(m))=I(f(\pi(m));h_\beta(\pi(m)))=I(f(m);h_\beta(m))=0$ for any subset $\beta \in \cB$ by the security of $({\mathcal F},\mathcal{G})$ on $\cI$.
\end{itemize}

\subsection*{{\bf Proof of Theorem \ref{thm:Bzero}:} 
Let $\cI = (G,S,D,\cB)$ be an instance of the key-dissemination problem with $\cB=\phi$, and let $\cI_\rs=(G,(S_m,S_r),D,\cB)$ be the corresponding instance of the secure multicast problem with $S_m=S_r=S$, then 
$$
\bRk^L(\cI)=\bRs^L(\cI_\rs).
$$
}

\proof
The fact that $\bRk^L(\cI) \geq \bRs^L(\cI_\rs)$ follows from our definitions as discussed above. 
To show that 
$\bRk^L(\cI) \leq \bRs^L(\cI_\rs)$, consider a linear network  code $({\mathcal F},\mathcal{G})=(\{f_{e}\},\{g_j\})$ for $\cI$ that is $(R,n)_\rk$-feasible. Let $K=Am$ where for $S=(s_i: i\in |S|)$, $m_i = (b_{ij}:j \in [\ell_i])$ are the independent random bits used by source $s_i$ in the communication process, $m=(b_{ij}:s_i \in S, j \in [\ell_i])$ is the vector of random bits used by all sources during communication, $\ell = \sum_{s_i \in S}{\ell_i}$ is the size of $m$, and $A$ is the $nR \times \ell$ global-encoding matrix of $K$.
Similar to the proof of Theorem~\ref{thm:single}, 
if $|m|=\ell=Rn$, then by our definitions it follows that $K$ must equal a linear permutation of $m$ and thus slightly modifying the decoding functions in $\cI_\rs$ to output $m$ we obtain an $(R,n)_\rs$-feasible  code for $\cI_\rs$.

Assume that $|m|>Rn$, we now claim that there exists $s_i\in S$ and $j \in [\ell_i]$ such that the matrix $A'$ obtained from $A$ by replacing the column in $A$ corresponding to $b_{ij}$ by the all zero column, satisfies $H(A'm)=Rn$.
This implies that a new key $K'=A'm$ of the same rate can be communicated using the same linear network  code $({\mathcal F},\mathcal{G})$ in which source $s_i$ replaces the random bit $b_{ij}$ by a constant value of 0, or equivalently, source $s_i$ omits random bit $b_{ij}$ from the linear combinations transmitted on its outgoing links.
The latter, in turn, implies that the modified code uses fewer bits from $m$, i.e., only $(\ell-1)$ bits instead of the previous $\ell$.
Continuing in this manner inductively, i.e., reducing the number of bits used from $m$ by zeroing out columns of $A$, we eventually obtain a linear multicast code for which exactly $Rn$ bits from $m$ are used to determine the uniform rate-$R$ key  shared by the terminals. 
This now implies, as discussed in the case that $|m|=Rn$, that $\cI_\rs$ is $(R,n)_\rs$-feasible.
Notice, that it is crucial that we are studying the case of $\cB=\phi$, as the process above does not necessarily preserve independence between the resulting key and other forms of information transmitted on network links. 

To prove the claim above, assume $|m|=\ell>Rn$. 
%This implies that $A$'s row rank (which is $Rn$) is strictly smaller than it's column dimension $\ell$. 
Thus there exists $s_i\in S$, $j \in [\ell_i]$ such that the column of $A$ corresponding to $b_{ij}$ is a linear combination of the remaining columns of $A$.
Let $A'$ be the matrix obtained from $A$ by zeroing out the column corresponding to $b_{ij}$.
By our construction, the rank of $A'$ equals that of $A$, or equivalently $H(A'm)=H(K')=Rn$.

\begin{remark}
	It is still open whether Theorem~\ref{thm:Bzero} holds for general (not necessarily linear) codes. 
	In other words, the answer to Question~\ref{q:mix} restricted to the non-secure setting, which asks if ``mixing helps'', is unknown.
	An affirmative answer would imply that $\bRk(\cI)>\bRs(\cI_\rs)$ and thus, as $\bRs(\cI_\rs) = \bRs^L(\cI_\rs)=\bRk^L(\cI)$ when $\cB=\phi$, that $\bRk(\cI) > \bRk^L(\cI)$, i.e., that there is an advantage to non-linear codes in key-dissemination when $\cB=\phi$.
\end{remark}

\subsection*{{\bf Proof of Theorem~\ref{thm:hard_k_s}:}
Let $\cI_\rs=(G,(S_m,S_r),D,\cB)$ be a secure-multicast instance with $|S_m|=1$.
%, in which 
%$G=(V,E)$, 
%$S_m=\{s\} \subseteq V$.
%, and $S_r=V$.
%and $\cB = \{\beta_e=\{e\} | e \in E\}$ consists of all single-edge subsets of $E$.
Let $R$ be a rate parameter.
One can efficiently construct an instance $\cI_\rk=(G_\rk,S_\rk,D_\rk,\cB_\rk)$  of the key dissemination problem such that 
$R \in \bRs(\cI_\rs)$ if and only if $R \in \bRk(\cI_\rk)$.}

%In \ml{\cite{zzz}}, it is shown that even for secure-multicast instances $\cI_\rs$ for which $S_m$ is of size 1, $D$ is of size 1, $\cB = \{\beta_e=\{e\} | e \in E\}$ consists of all single-edge subsets of $E$, and $S_r=V$, computing the secure-multicast capacity is as hard as resolving the capacity of multiple-unicast network coding instances. 
%We this conclude the following corollary for the resulting $\cI_\rk$ obtained in the reduction of from Theorem~\ref{thm:hard_k_s}:
%\begin{corollary}
%\label{cor:hard}
%Determining the capacity of the key-dissemination problem is at least as hard as determining the capacity of the multiple-unicast network coding problem.
%\end{corollary}
%\ml{Need to verify the exact hardness statement}
%Computing $\bRk$ is as hard as computing $\bRs$.
%The latter, in turn is as hard as computing the multiple-unicast network coding capacity \cite{}.}

\proof 
%(of Theorem~\ref{thm:hard_k_s})
Let $R$ be a given rate parameter.
%Let $\cI=(G,s,D))$ be a secure network coding instance with $G=(V,E)$.
%We first modify $\cI$ slightly by {\em splitting} each edge $e=(u,v) \in E$ into two edges in $(u,w_{uv})$ and $(w_{uv},v)$.
%We denote the modified $\cI$ as $\cI^{\tt split}=(G^{\tt split},s,D)$.
%It is not hard to verify that $R \in \bRs(\cI)$ if and only if $R \in \bRs(\cI^{\tt split})$.
We first construct the instance $\cI_\rk=(G_\rk,S_\rk,D_\rk,\cB_\rk)$ of the key dissemination problem
with $G_\rk=({V}_\rk,{E}_\rk)$.
Instance $\cI_\rk$ is obtained from $\cI_\rs$ by adding a new terminal $d_\rk$ to the set $D$ to obtain $V_\rk = V \cup \{d_\rk\}$ and $D_\rk=D \cup \{d_\rk\}$, by adding a new edge of capacity $R$ connecting source $s$ of $S_m$ with $d_\rk$ to give $E_\rk=E \cup \{(s,d_\rk)\}$, by setting $\cB_\rk=\cB$, and by setting $S_\rk=S_m \cup S_r$. 
%Correspondingly,  $E_\rk=E \cup \{(s,d_\rk):s \in S_m\}$.
 
We show for every $R' \leq R$ that there exists an $(R', n)_\rs$-feasible code for $\cI_\rs$ if and only if there exists an $(R',n)_\rk$-feasible code for $\cI_\rk$.
First assume that there exists an $(R', n)_\rs$-feasible code for $\cI_\rs$. 
Let $K$ be the rate $R'$ message that is securely communicated from $S_m$ to all terminals in $D$. Using the exact same code on $\cI_\rk$ and communicating $K$ directly on the new edge $(s,d_\rk)$, one can communicate $K$ to all terminals in ${D}_\rk$. As the original code is secure in $\cI_\rs$ for the edge sets in $\cB=\cB_\rk$, the code is $(R',n)_\rk$-feasible for $\cI_\rk$.

Now assume that there exists an $(R',n)_\rk$-feasible code for $\cI_\rk$.
As only information generated at $S_m$ can be shared between the new terminal $d_\rk$ and other terminals in $D_\rk$, it holds that the shared uniform key $K$ is a function of the random bits generated at $s$ of $S_m$.
Moreover, as $\cB = \cB_\rk$, for every $\beta \in \cB$ the code on $\cI_\rk$ satisfies $I(K,(X^n_e: e \in \beta))=0$.
Now, using ideas of Theorem~\ref{thm:single}, one can pre-encode at $s$ to obtain a code that, when restricted to $G$, is an $(R', n)_\rs$-feasible code for $\cI_\rs$.
This concludes the proof of our assertion.

\subsection*{{\bf Proof of Theorem~\ref{thm:gap}:}
	For any integer $\alpha>1$, there exist instances $\cI = (G,S,D,\cB)$ of the key-dissemination problem such that   
$$
\bRk(\cI) \ge \alpha\bRkk(\cI)
$$
}

\proof
Let $\alpha >1$. Roughly speaking, the instance $\cI=(G,S,D,\cB)$ we present is reminiscent of the {\em combination network} \cite{ngai2004network}.  
The network $\cI$, depicted in a simplified form in Figure~\ref{fig:examples}.a for the special case of $\alpha = 2$, has the following structure.
$G$ is acyclic and has three layers of nodes. The first layer consists of the source nodes $S=\{s_1,\dots,s_r\}$. Here, we set $r$ to be equal to $\alpha+1$.
%take $r$ such that $r-1 \geq \alpha$. 
The second layer  consists of two sets of intermediate nodes $U=\{u_1,\dots,u_r\}$ and $\bar{U}=\{\bar{u}_1,\dots,\bar{u}_r\}$.
%The third layer duplicates the second one and consists of nodes  $W=\{w_1,\dots,w_r\}$ and $\bar{W}=\{\bar{w}_1,\dots,\bar{w}_r\}$.
The final layer consists of terminal nodes $D=\{d_{i}\}_{i \in [r]}$.
The edge set of $G$ consists of the following edges, an edge $(s_i,u_i)$ for every $i \in [r]$,  an edge $(s_j,\bar{u}_i)$ for every $j \ne i$ in $[r]^2$, an edge $(u_i,d_i)$ and $(\bar{u}_i,d_i)$ for every $i \in [r]$.
%, an edge $(w_i,d_{i})$ and an edge $(\bar{w}_i,d_{i})$ for every $i \in [r]$.
Each edge has capacity 1.
For each node $v \in U \cup \bar{U}$, the set $\cB$ contains a subset $\beta_v=(e: e \in {\tt In}(v))$  comprising all incoming edges to $v$. 
Thus, $\cB=\{\beta_v : v \in U \cup \bar{U}\}$.

We first show that $\bRk(\cI) \leq 1$. 
Consider any network  code  for $\cI$  that is $(R,n)_\rk$-feasible. 
Let $K$ be the key shared by all terminal nodes.
For nodes $v \in U \cup \bar{U}$, let $e(v)$ be the (single) edge leaving $v$ and let $X_{e(v)}^n$ be the information transmitted on $e(v)$.
Since $\beta_v \in \cB$, it must hold that $I(K;X_{e(v)}^n) \leq I(K;(X_e^n : e \in \beta_v))=0$.
We now show that this implies that $R \leq 1$.
Let $i \in [r]$, and consider terminal $d_i$. 
The structure of $\cI$ implies that 
\begin{align*}
H(K) & = I(K;X_{e(u_i)}^n,X_{e(\bar{u}_i)}^n) \\
& = I(K;X_{e(u_i)}^n)+I(K;X_{e(\bar{u}_i)}^n|X_{e(u_i)}^n)) \\
& = I(K;X_{e(\bar{u}_i)}^n|X_{e(u_i)}^n)) \leq H(X_{e(\bar{u}_i)}^n) \leq n.
\end{align*}
To show that $\bRk(\cI)=1$, we present a  network  code for $\cI$  that is $(1,n)_\rk$-feasible (i.e., of rate $R=1$). 
Roughly speaking, our code communicates  the sum of all sources to each terminal $d_i$.
Formally, for $n=1$, source node $s_i$ sends a single bit $b_{i}$ on all its outgoing edges, and nodes $u_i$ and $\bar{u_i}$ send the binary sum of their incoming information on their single outgoing edge. Summing these, every terminal obtains the (shared) sum $\sum_{i=1}^r b_{i}$. Due to the nature of $K$, for any $\beta_v \in \cB$ it holds that $I(K;(X^n_{e}: e \in \beta_v))=0$. We conclude that $\cI$  is $(R,n)_\rk$-feasible for $R=1$.

\balance

We now show that $\bRkk(\cI) \leq \frac{1}{r-1}$.
Consider any network  code  for $\cI$  that is $(R,n)_{\rk(2)}$-feasible. 
Let the decoded messages from the first decoding stage be $M=(b_{ij}: (i,j) \in I)$ and let $K$ be the key obtained by the second stage.
Recall that $H(K|M)=0$.
Let $M_i = (b_{ij}: (i,j) \in I)$ be the bits in $M$ generated at source $s_i \in S$, and let $R_i = M_i/n$. 
For any $i \in [r]$, removing a single edge from $\cI$ separates terminal $d_i$ from sources $(s_j: j \ne i)$.
Therefore, using standard cut-set bounds with respect to terminal $d_i$, it holds that
$\sum_{j \ne i}{R_i} \leq 1$.
By summing the above over $i$, we conclude that 
$\sum_i\sum_{j \ne i}{R_i} \leq r$, which in turn implies that 
$\sum_i{R_i} \leq \frac{r}{r-1}$.
Moreover, as $u_i \in U$ lies on the only path from $s_i$ to terminal $d_i$, $H(M_i|X^n_{{\tt In}(u_i)})=0$.
By our definition of $\cB$ we have for all $i \in [r]$ that $I(K;X^n_{{\tt In}(u_i)})=0$, which now implies that $I(K;M_i)=0$  for all $i \in [r]$.
Similarly, by our definition of $\cB$, it holds that $I(K;(M_j: j\ne i))=0$ for all $i \in [r]$ since $\bar{u}_i \in \bar{U}$ lies on the only path from $\{s_j\}_{j \ne i}$ to $d_i$.
Thus, for every $i \in [r]$,
\begin{align*}
H(K) & =I(K;M) \\
& =I(K;(M_j: j\ne i)) + I(K;M_i|(M_j: j\ne i)) \\
& =  I(K;M_i|(M_j: j\ne i)) \leq H(M_i) = R_in.
\end{align*}
Summing over all $i \in [r]$, we therefor conclude that $rH(K) \leq n\sum_i{R_i} \leq n \cdot \frac{r}{r-1}$, implying that 
$Rn=H(K) \leq \frac{n}{r-1}$.
We conclude that 
$$
1=\bRk(\cI) \geq (r-1)\bRkk(\cI) = \alpha \bRkk(\cI).
$$

%the instances used in Theorem~\ref{thm:gap} are generalized variants of {\em combination networks} \cite{ngai2004network}, that, on one hand, allow a key capacity of 1 by multicasting a key $K$ equal to the sum-of-sources, and, on the other, are designed to have a limited non-secure multicast sum-rate. The later, together with the pre-defined security requirements, limits the 2-stage key-capacity to obtain the stated gap. 
%A simplified example is depicted in Figure~\ref{fig:examples}.a for the special case of $\alpha = 2$.

\section{Conclusions}
\label{sec:conclude}

This work addresses the key-dissemination problem in the context of network coding, in which
a number of results comparing key capacity with the traditional secure-multicast capacity are presented. 
For single-source networks and linear non-secure networks, we show that there is no rate advantage in the flexible nature of the shared key $K$ in key-dissemination when compared to the requirement of secure-multicast that $K$ include source information bits. For general instances, we demonstrate rate advantages of key-dissemination when compared to secure-multicast or restricted forms of 2-stage key-dissemination decoding. 
Finally,  we show that determining the key capacity is as hard as determining the secure-multicast capacity which, in turn, is as hard as determining the multiple-unicast network coding capacity.

Several questions remain open or unstudied in this work. 
For the non-secure (multiple-source) setting, it is currently unresolved whether {\em mixing} 
(in the sense of Question~\ref{q:mix}) allows improved key rates compared to traditional multi-source multicast. This work does not address the multiple-multicast analog of key-dissemination in which different sets of terminals require independent secret keys, potentially mutually hidden between the different terminal sets.  Understanding the multiple-multicast analog of key-dissemination exhibits challenges even for the 2-multicast case and has strong connections to the cryptographic study of {\em secret sharing}. Finally, efficient communication schemes, especially designed for the multicast (or the multiple-multicast analog) of key-dissemination are not presented in this work. 
While one can design multicast key-dissemination schemes relying on random linear network coding enhanced with certain security measures, a comprehensive study in this aspect is the subject of ongoing work.

%\bibliographystyle{IEEEtran}
%\newpage
\thispagestyle{empty}

%{\tiny{
\bibliographystyle{unsrt}
\bibliography{ref}

\begin{thebibliography}{10}

\bibitem{wyner1975wire}
Aaron~D Wyner.
\newblock The wire-tap channel.
\newblock {\em Bell system technical journal}, 54(8):1355--1387, 1975.

\bibitem{csiszar1978broadcast}
Imre Csisz{\'a}r and Janos Korner.
\newblock Broadcast channels with confidential messages.
\newblock {\em IEEE transactions on information theory}, 24(3):339--348, 1978.

\bibitem{ahlswede1993common}
Rudolf Ahlswede and Imre Csisz{\'a}r.
\newblock {Common randomness in information theory and cryptography. I. Secret
  sharing}.
\newblock {\em IEEE Transactions on Information Theory}, 39(4):1121--1132,
  1993.

\bibitem{maurer1993secret}
Ueli~M Maurer.
\newblock Secret key agreement by public discussion from common information.
\newblock {\em IEEE transactions on information theory}, 39(3):733--742, 1993.

\bibitem{csiszar2004secrecy}
Imre Csisz{\'a}r and Prakash Narayan.
\newblock Secrecy capacities for multiple terminals.
\newblock {\em IEEE Transactions on Information Theory}, 50(12):3047--3061,
  2004.

\bibitem{chan2014multiterminal}
Chung Chan and Lizhong Zheng.
\newblock Multiterminal secret key agreement.
\newblock {\em IEEE transactions on information theory}, 60(6):3379--3412,
  2014.

\bibitem{csiszar2008secrecy}
Imre Csisz{\'a}r and Prakash Narayan.
\newblock Secrecy capacities for multiterminal channel models.
\newblock {\em IEEE Transactions on Information Theory}, 54(6):2437--2452,
  2008.

\bibitem{gohari2010information}
Amin~Aminzadeh Gohari and Venkat Anantharam.
\newblock {Information-theoretic key agreement of multiple terminals—Part I}.
\newblock {\em IEEE Transactions on Information Theory}, 56(8):3973--3996,
  2010.

\bibitem{gohari2010information2}
Amin~Aminzadeh Gohari and Venkat Anantharam.
\newblock {Information-theoretic key agreement of multiple terminals—Part II:
  Channel model}.
\newblock {\em IEEE Transactions on Information Theory}, 56(8):3997--4010,
  2010.

\bibitem{siavoshani2010group}
Mahdi~Jafari Siavoshani, Christina Fragouli, Suhas Diggavi, Uday Pulleti, and
  Katerina Argyraki.
\newblock Group secret key generation over broadcast erasure channels.
\newblock In {\em Forty Fourth IEEE Asilomar Conference on Signals, Systems and
  Computers}, pages 719--723, 2010.

\bibitem{xu2016private}
Peng Xu, Zhiguo Ding, Xuchu Dai, and George~K. Karagiannidis.
\newblock On the private key capacity of the $ m $-relay pairwise independent
  network.
\newblock {\em IEEE Transactions on Information Theory}, 62(7):3831--3843,
  2016.

\bibitem{hayashi2016secret}
Masahito Hayashi, Himanshu Tyagi, and Shun Watanabe.
\newblock Secret key agreement: General capacity and second-order asymptotics.
\newblock {\em IEEE Transactions on Information Theory}, 62(7):3796--3810,
  2016.

\bibitem{narayan2016multiterminal}
Prakash Narayan and Himanshu Tyagi.
\newblock {\em Multiterminal secrecy by public discussion}.
\newblock Now Publishers Hanover, MA, USA, 2016.

\bibitem{cai2002secure}
Ning Cai and Raymond~W Yeung.
\newblock Secure network coding.
\newblock {\em IEEE International Symposium on Information Theory}, page 323,
  2002.

\bibitem{feldman2004capacity}
Jon Feldman, Tal Malkin, C~Stein, and RA~Servedio.
\newblock On the capacity of secure network coding.
\newblock {\em 42nd Annual Allerton Conference on Communication, Control, and
  Computing}, pages 63--68, 2004.

\bibitem{cai2007security}
Ning Cai and Raymond~W Yeung.
\newblock A security condition for multi-source linear network coding.
\newblock {\em IEEE International Symposium on Information Theory}, pages
  561--565, 2007.

\bibitem{yeung2008optimality}
Ning Cai and Raymond~W Yeung.
\newblock On the optimality of a construction of secure network codes.
\newblock {\em IEEE International Symposium on Information Theory}, pages
  166--170, 2008.

\bibitem{el2012secure}
Salim~El Rouayheb, Emina Soljanin, and Alex Sprintson.
\newblock {Secure network coding for wiretap networks of type II}.
\newblock {\em IEEE Transactions on Information Theory}, 58(3):1361--1371,
  2012.

\bibitem{silva2011universal}
Danilo Silva and Frank~R Kschischang.
\newblock Universal secure network coding via rank-metric codes.
\newblock {\em IEEE Transactions on Information Theory}, 57(2):1124--1135,
  2011.

\bibitem{jaggi2012secure}
Sid Jaggi and Michael Langberg.
\newblock Secure network coding: Bounds and algorithms for secret and reliable
  communications.
\newblock In {\em Chapter 7 of Network Coding: Fundamentals and applications
  (Muriel M{\'e}dard and Alex Sprintson ed.)}, pages 183--215. Academic Press,
  2012.

\bibitem{huang2018}
W.~Huang, T.~Ho, M.~Langberg, and J.~Kliewer.
\newblock Single-unicast secure network coding and network error correction are
  as hard as multiple-unicast network coding.
\newblock {\em IEEE Transactions on Information Theory}, 64(6):4496--4512,
  2018.

\bibitem{chan2014network}
Terence~H Chan and Alex Grant.
\newblock Network coding capacity regions via entropy functions.
\newblock {\em IEEE Transactions on Information Theory}, 60(9):5347--5374,
  2014.

\bibitem{cui2012secure}
Tao Cui, Tracy Ho, and Joerg Kliewer.
\newblock On secure network coding with nonuniform or restricted wiretap sets.
\newblock {\em IEEE Transactions on Information Theory}, 59(1):166--176, 2012.

\bibitem{chaudhuri2018trade}
Debaditya Chaudhuri and Michael Langberg.
\newblock Trade-offs between rate and security in linear multicast network
  coding.
\newblock In {\em IEEE International Symposium on Information Theory (ISIT)},
  pages 846--850, 2018.

\bibitem{chaudhuri2019secure}
Debaditya Chaudhuri, Michael Langberg, and Michelle Effros.
\newblock Secure network coding in the setting in which a non-source node may
  generate random keys.
\newblock In {\em IEEE International Symposium on Information Theory (ISIT)},
  pages 2309--2313, 2019.

\bibitem{chaudhuri2021characterization}
Debaditya Chaudhuri.
\newblock {\em Characterization of Rate Regions in Secure Network Coding over
  General Wiretap Networks}.
\newblock PhD thesis, University at Buffalo, State University of New York,
  2021.

\bibitem{4460828}
T.~Chan and A.~Grant.
\newblock Capacity bounds for secure network coding.
\newblock {\em Australian Communications Theory Workshop}, pages 95--100, 2008.

\bibitem{6293890}
T.~Cui, T.~Ho, and J.~Kliewer.
\newblock On secure network coding with nonuniform or restricted wiretap sets.
\newblock {\em IEEE Transactions on Information Theory}, 59(1):166--176, 2013.

\bibitem{langberg2009multiple}
Michael Langberg and Muriel M{\'e}dard.
\newblock On the multiple unicast network coding, conjecture.
\newblock {\em 47th Annual Allerton Conference on Communication, Control, and
  Computing}, pages 222--227, 2009.

\bibitem{appuswamy2011network}
Rathinakumar Appuswamy, Massimo Franceschetti, Nikhil Karamchandani, and
  Kenneth Zeger.
\newblock Network coding for computing: Cut-set bounds.
\newblock {\em IEEE Transactions on Information Theory}, 57(2):1015--1030,
  2011.

\bibitem{kowshik2012optimal}
Hemant Kowshik and PR~Kumar.
\newblock Optimal function computation in directed and undirected graphs.
\newblock {\em IEEE Transactions on Information Theory}, 58(6):3407--3418,
  2012.

\bibitem{shah2013network}
Virag Shah, Bikash~Kumar Dey, and D~Manjunath.
\newblock Network flows for function computation.
\newblock {\em IEEE Journal on Selected Areas in Communications},
  31(4):714--730, 2013.

\bibitem{ramamoorthy2013communicating}
Aditya Ramamoorthy and Michael Langberg.
\newblock Communicating the sum of sources over a network.
\newblock {\em IEEE Journal on Selected Areas in Communications},
  31(4):655--665, 2013.

\bibitem{rai2012network}
Brijesh~Kumar Rai and Bikash~Kumar Dey.
\newblock On network coding for sum-networks.
\newblock {\em IEEE Transactions on Information Theory}, 58(1):50--63, 2012.

\bibitem{shenvi2010necessary}
Sagar Shenvi and Bikash~Kumar Dey.
\newblock A necessary and sufficient condition for solvability of a 3s/3t
  sum-network.
\newblock In {\em 2010 IEEE International Symposium on Information Theory},
  pages 1858--1862, 2010.

\bibitem{appuswamy2013computing}
Rathinakumar Appuswamy and Massimo Franceschetti.
\newblock Computing linear functions by linear coding over networks.
\newblock {\em IEEE transactions on information theory}, 60(1):422--431, 2013.

\bibitem{li2022arithmetic}
Sijie Li and Cheuk~Ting Li.
\newblock Arithmetic network coding for secret sum computation.
\newblock {\em arXiv preprint arXiv:2201.03032}, 2022.

\bibitem{brahma2015pliable}
Siddhartha Brahma and Christina Fragouli.
\newblock Pliable index coding.
\newblock {\em IEEE Transactions on Information Theory}, 61(11):6192--6203,
  2015.

\bibitem{bar2011index}
Ziv Bar-Yossef, Yitzhak Birk, TS~Jayram, and Tomer Kol.
\newblock Index coding with side information.
\newblock {\em IEEE Transactions on Information Theory}, 57(3):1479--1494,
  2011.

\bibitem{el2010index}
Salim El~Rouayheb, Alex Sprintson, and Costas Georghiades.
\newblock On the index coding problem and its relation to network coding and
  matroid theory.
\newblock {\em IEEE Transactions on Information Theory}, 56(7):3187--3195,
  2010.

\bibitem{effros2015equivalence}
Michelle Effros, Salim El~Rouayheb, and Michael Langberg.
\newblock An equivalence between network coding and index coding.
\newblock {\em IEEE Transactions on Information Theory}, 61(5):2478--2487,
  2015.

\bibitem{dau2012security}
Son~Hoang Dau, Vitaly Skachek, and Yeow~Meng Chee.
\newblock On the security of index coding with side information.
\newblock {\em IEEE Transactions on Information Theory}, 58(6):3975--3988,
  2012.

\bibitem{liu2019private}
Tang Liu and Daniela Tuninetti.
\newblock Private pliable index coding.
\newblock In {\em IEEE Information Theory Workshop (ITW)}, pages 1--5, 2019.

\bibitem{sasi2019code}
Shanuja Sasi and B~Sundar Rajan.
\newblock Code construction for pliable index coding.
\newblock In {\em IEEE International Symposium on Information Theory (ISIT)},
  pages 527--531, 2019.

\bibitem{liu2020secure}
Tang Liu and Daniela Tuninetti.
\newblock Secure decentralized pliable index coding.
\newblock In {\em IEEE International Symposium on Information Theory (ISIT)},
  pages 1729--1734, 2020.

\bibitem{oliveira2008network}
Paulo~F Oliveira and Joao Barros.
\newblock A network coding approach to secret key distribution.
\newblock {\em IEEE Transactions on Information Forensics and Security},
  3(3):414--423, 2008.

\bibitem{xiao2018cooperative}
Shuaifang Xiao, Yunfei Guo, Kaizhi Huang, and Liang Jin.
\newblock Cooperative group secret key generation based on secure network
  coding.
\newblock {\em IEEE Communications Letters}, 22(7):1466--1469, 2018.

\bibitem{lima2009towards}
Luisa Lima, Joao Barros, Muriel M{\'e}dard, and Alberto Toledo.
\newblock Towards secure multiresolution network coding.
\newblock In {\em IEEE Information Theory Workshop on Networking and
  Information Theory}, pages 125--129, 2009.

\bibitem{langberg2021edge}
Michael Langberg and Michelle Effros.
\newblock Edge removal in undirected networks.
\newblock In {\em IEEE International Symposium on Information Theory (ISIT)},
  pages 1421--1426, 2021.

\bibitem{ngai2004network}
Chi~Kin Ngai and Raymond~W Yeung.
\newblock Network coding gain of combination networks.
\newblock In {\em Information Theory Workshop}, pages 283--287. IEEE, 2004.

\end{thebibliography}
%}}

\end{document}